\documentclass[12pt]{article}
\usepackage{amsfonts,amsmath}
\usepackage{epsfig,graphics}

\renewcommand{\appendix}{%
\renewcommand{\section}{%
\newpage
\thispagestyle{plain}%
\secdef\Appendix\sAppendix}%
\setcounter{section}{0}%
\renewcommand{\thesection}{\Alph{section}}%
}
\newcommand{\Appendix}[2][?]{%
\refstepcounter{section}%
\addcontentsline{toc}{Addendum}%
{\protect\numberline{\appendixname~\thesection}#1}%
{\flushleft\LARGE\bfseries\appendixname\ \thesection\par
\centering#2\par}%
\sectionmark{#1}\vspace{\baselineskip}}

\newcommand{\sAppendix}[1]{%
{\flushright\large\bfseries\appendixname\par
\centering#1\par}%
\vspace{\baselineskip}}
\def\be{\begin{equation}}

\def\bea{\begin{eqnarray}}

\def\eea{\end{eqnarray}}

\normalbaselines \oddsidemargin 0.5cm \evensidemargin 0.5cm
\begin{document}

\pagestyle{empty}


\vskip 1cm

\begin{center}

{\LARGE {\textbf{Unitary Braid Matrices: Bridge between
Topological and Quantum Entanglements }}} \vspace{4mm}

{\bf \large B. Abdesselam$^{a,}$\footnote{Email:
boucif@cpht.polytechnique.fr and  boucif@yahoo.fr} and A.
Chakrabarti$^{b,}$\footnote{Email: chakra@cpht.polytechnique.fr}}

\vspace{2mm}

\emph{$^a$ Laboratoire de Physique Th\'eorique, Universit\'e
d'Oran Es-S\'enia, 31100-Oran, Alg\'erie\\
and \\ Facult\'e des Sciences et de la Technologie, Centre Universitaire d'Ain-T\'emouchent, 46000-Ain-T\'emouchent, Alg\'erie}\\
\vspace{2mm}\emph{$^b$ Centre de Physique Th{\'e}orique, Ecole
Polytechnique, 91128 Palaiseau Cedex, France.}
\end{center}

\begin{abstract}
{\small \noindent Braiding operators corresponding to the third
Reidemeister move in the theory of knots and links are realized in
terms of parametrized unitary matrices for all dimensions. Two
distinct classes are considered. Their (non-local) unitary actions
on separable pure product states of three identical subsystems
(say, the spin projections of three particles) are explicitly
evaluated for all dimensions. This, for our classes, is shown to
generate entangled superposition of four terms in the base space.
The 3-body and 2-body entanglements (in three 2-body subsystems),
the 3-tangles and 2-tangles are explicitly evaluated for each
class. For our matrices, these are parametrized. Varying
parameters they can be made to sweep over the domain (0,1).Thus
braiding operators corresponding to over- and under-crossings of
three braids and, on closing ends, to topologically entangled
Borromean rings are shown, in another context, to generate quantum
entanglements. For higher dimensions, starting with different
initial triplets one can entangle by turns, each state with all
the rest. A specific coupling of three angular momenta is briefly
 discussed to throw more light on three body entanglements.}
\end{abstract}

\newpage

\pagestyle{plain} \setcounter{page}{1}

\section{Introduction (two faces of unitary braid matrices)}
\setcounter{equation}{0}

The third Reidemeister move in the theory of knots and links
impose equivalence between two specific sequences of over- and
under-crossing of three braids. Such sequences can be repeated and
closing the ends of the braids one obtains topologically entangled
Borromean rings whose history reaches back far into the past (see
fig. 7 of Ref. 1). Braid matrices, "Baxterized" to depend on
spectral (rapidity) parameters satisfy an equation - the braid
equation - which corresponds precisely to the above mentioned
Reidemeister move. They provide matricial representations of a
particular type of topological entanglements. When a braid matrix
is also unitary it can also be implemented to induce unitary
transformations in base spaces of corresponding dimensions
representing possible quantum states of an object (say, the spin
projections of particles).

Let us make this more precise. Let
$\widehat{\mathrm{R}}\left(\theta\right)$ be a unitary $N^2\times
N^2$ matrix, I the $N\times N$ unit matrix and
\begin{equation}
\widehat{\mathrm{R}}_{12}=\widehat{\mathrm{R}}\otimes I,\qquad
\widehat{\mathrm{R}}_{23}=I\otimes
\widehat{\mathrm{R}},\end{equation} $\widehat{\mathrm{R}}_{12}$,
$\widehat{\mathrm{R}}_{23}$ act on triple tensor products
$\mathrm{V}_N\otimes \mathrm{\mathrm{V}}_N\otimes \mathrm{V}_N$ of
$N$-dimensional vector spaces $\mathrm{V}_N$. To be a braid matrix
$\widehat{\mathrm{R}}\left(\theta\right)$ must satisfy
\begin{equation}
\widehat{\mathrm{R}}_{12}\left(\theta\right)\widehat{\mathrm{R}}_{23}\left(\theta+\theta'\right)\widehat{\mathrm{R}}_{12}\left(\theta'\right)=
\widehat{\mathrm{R}}_{23}\left(\theta'\right)\widehat{\mathrm{R}}_{12}\left(\theta+\theta'\right)\widehat{\mathrm{R}}_{23}\left(\theta\right).
\end{equation}
The indices $(12)$, $(23)$ correspond to successive crossings
(braid 2 overcrossing braid 1 and undercrossing braid 3). The
equality sign imposes the essential Reidemeister constraint (the
3-rd move). This will be the link of our matrices to topological
entanglement. The role of braid matrices satisfying unitarity
\begin{equation}
\left(\widehat{\mathrm{R}}\left(\theta\right)\right)^+=\widehat{\mathrm{R}}\left(\theta\right)
\end{equation}
in quantum entanglements have been noted and discussed by Kauffman
and Lomonaco \cite{1} in their paper "Braiding operators are
universal quantum gates". A large number of relevant sources are
cited in Ref 1. We have presented before two quite distinct
classes of unitary $N^2\times N^2$ braid matrices \cite{2,3}, one
real and for even $N$ and the other complex, for All $N$. There is
no upper limit to $N$. Refs. 2, 3 cite other sources.

In the following sections we will systematically, explicitly
derive the measures of 2-body and 3-body entanglements generated
by the action of the braiding operator
\begin{equation}
\widehat{\mathrm{B}}=\widehat{\mathrm{R}}_{12}\left(\theta\right)\widehat{\mathrm{R}}_{23}\left(\theta+\theta'\right)\widehat{\mathrm{R}}_{12}
\left(\theta'\right)\left(=
\widehat{\mathrm{R}}_{23}\left(\theta'\right)\widehat{\mathrm{R}}_{12}\left(\theta+\theta'\right)\widehat{\mathrm{R}}_{23}\left(\theta\right)\right)
\end{equation}
acting on the pure separable product states
\begin{equation}
\left|a\right\rangle\otimes\left|b\right\rangle\otimes\left|c\right\rangle=\left|abc\right\rangle
\end{equation}
spanning the basis $\mathrm{V}_N\otimes \mathrm{V}_N\otimes
\mathrm{V}_N$. One can then, if necessary, evaluate
\begin{equation}
\widehat{\mathrm{B}}\left(\sum_{a,b,c}f_{abc}\left|abc\right\rangle\right)
\end{equation}
The measures of 2-tangles and 3-tangles derived in Ref. 4 will be
used throughout. Topological and quantum entanglements, two
domains of $\widehat{\mathrm{B}}$, will thus be brought together.
One essential point must be noted: The unitary matrix
$\widehat{\mathrm{R}}$ is not locally unitary.
$\widehat{\mathrm{R}}$ cannot be expressed as
$\widehat{\mathrm{R}}_1\otimes \widehat{\mathrm{R}}_2$ acting on
$\mathrm{V}_N\otimes \mathrm{V}_N$ where $\widehat{\mathrm{R}}_1$,
$\widehat{\mathrm{R}}_2$ are each a unitary $N\times N$ matrix
acting on $\mathrm{V}_N$. Such an $\widehat{\mathrm{R}}$ would
have been trivial in the context of braiding. Nor would they have
induced quantum entanglements acting on a product state
$\left|a\right\rangle\otimes\left|b\right\rangle$ in
$\mathrm{V}\otimes \mathrm{V}$. Non-local unitarity is crucial in
the action of $\widehat{\mathrm{B}}$. Local unitary
transformations can however be used to classify already entangled
states, as has been done systematically by Carteret and Sulbery
\cite{5}. We aim at generating quantum entanglements.

We close the introduction with some notations we will use
throughout. For $N=2$, $\mathrm{V}_2$ is usually taken to be
spanned by the spin projections of spin-$\frac 12$ particles
\begin{equation}
\left(\left|+\right\rangle,\left|-\right\rangle\right)\equiv
\left(\left|1\right\rangle,\left|0\right\rangle\right)
\end{equation}
With passage to higher dimensions in mind we will often use the
state vectors
\begin{equation}
\left|1\right\rangle=\left|\begin{array}{c} 1 \\0 \\
\end{array}\right\rangle,
\qquad
\left|\overline{1}\right\rangle=\left|\begin{array}{c} 0 \\1 \\
\end{array}\right\rangle
\end{equation}
Generalization is direct. Thus for $N=4$,
\begin{equation}
\left|1\right\rangle=\left|\begin{array}{c} 1 \\0 \\0 \\0 \\
\end{array}\right\rangle,
\qquad
\left|2\right\rangle=\left|\begin{array}{c} 0 \\1 \\0 \\0 \\
\end{array}\right\rangle,
\qquad
\left|\overline{2}\right\rangle=\left|\begin{array}{c} 0 \\0 \\1 \\0 \\
\end{array}\right\rangle,
\qquad
\left|\overline{1}\right\rangle=\left|\begin{array}{c} 0 \\0 \\0 \\1 \\
\end{array}\right\rangle.
\end{equation}
For all $N$,
\begin{equation}
\overline{i}=N+1-i,\qquad
\overline{\overline{i}}=N+1-\overline{i}=i.\end{equation} For
convenience we will continue to use the terminology of spin
projections. But the indices above can also correspond to other
suitably enumerated quantum states of a system.

\section{Unitary braid matrices and their actions}
\setcounter{equation}{0}

\paragraph{(I) Real, unitary, even-dimensional braid matrices:}
A class of real, unitary, $\left(2n\right)^2 \times
\left(2n\right)^2$ dimensional braid matrices \cite{2} is given by
\begin{equation}
\left(\widehat{\mathrm{R}}\left(z\right)\right)^{\pm 1}=\frac
1{\sqrt{1+z^2}}\left(I\otimes I\pm zK\otimes
J\right),\end{equation} where
\begin{equation}
z=\tanh\theta\end{equation} and $\left((K,J\right)$ are
$\left(2n\right)\times\left(2n\right)$ matrices given by
\begin{equation}
J=\sum_{i=1}^n\left((-1)^{\overline{i}}\left(i\overline{i}\right)+
(-1)^{i}\left(\overline{i}i\right)\right),\qquad
K=\sum_{i=1}^n\left(\left(i\overline{i}\right)+
\left(\overline{i}i\right)\right)\end{equation} with
$\overline{i}=2n+1-i$ and $I$ the $\left(2n\right)\times
\left(2n\right)$ unit matrix. We always denote by
$\left(ij\right)$ the matrix with a single non-zero element, unity
on row $i$ and column $j$. A detailed study of this class can be
found in Ref. 2. An equivalent construction, without explicit
introduction of the tensor product structure $\left(K\otimes
J\right)$, can be found in Ref. 6. From (2.3),
\begin{eqnarray}
&&JK=-KJ=\sum_{i=1}^n\left((-1)^{\overline{i}}\left(ii\right)+
(-1)^{i}\left(\overline{i}\overline{i}\right)\right),\nonumber\\
&&K^2=-J^2=I.
\end{eqnarray}
Denoting
$\left(\tanh\theta,\tanh\theta',\tanh\left(\theta+\theta'\right)\right)=
\left(z,z',z''\right)$ with
\begin{equation}z''=\frac{z+z'}{1+zz'}\end{equation}
and using (2.4) one obtains unitarity
\begin{equation}
\left(\widehat{\mathrm{R}}\left(z\right)\right)^+=\widehat{\mathrm{R}}\left(z\right)^{-1}
\end{equation}
and the explicit evaluation
\begin{eqnarray}
&&\widehat{\mathrm{B}}=\widehat{\mathrm{R}}_{12}\left(z\right)\widehat{\mathrm{R}}_{23}\left(z''\right)\widehat{\mathrm{R}}_{12}\left(z'\right)=
\widehat{\mathrm{R}}_{23}\left(z'\right)\widehat{\mathrm{R}}_{12}\left(z''\right)\widehat{\mathrm{R}}_{23}\left(z\right)\nonumber\\
&&\phantom{\widehat{\mathrm{B}}}=\frac
1{\sqrt{\left(1+z^2\right)\left(1+z'^2\right)\left(1+z''^2\right)}}\left(\left(1-zz'\right)I\otimes
I\otimes I+\right.\nonumber\\
&&\phantom{\widehat{\mathrm{B}}=}\left.\left(z+z'\right)\left(I\otimes
K\otimes J+K\otimes J\otimes
I\right)+z''\left(z'-z\right)\left(K\otimes KJ\otimes
J\right)\right).\end{eqnarray} (A misprint in the overall factor
in (2.16) of Ref. 2 is corrected above. We have also set
$z''\left(1+zz'\right)=\left(z+z'\right)$ in the second term
there.) We now consider the action of $\widehat{\mathrm{B}}$ on
basis states
\begin{equation}
\left|abc\right\rangle=\left|a\right\rangle\otimes\left|b\right\rangle\otimes\left|c\right\rangle,
\end{equation}
where $\left(a,b,c\right)$ each span the $\left(2n\right)$
dimensional $\mathrm{V}_{2n}$ in $\mathrm{V}_{2n}\otimes
\mathrm{V}_{2n}\otimes \mathrm{\mathrm{V}}_{2n}$ with for
\begin{eqnarray}
&&a=i,\qquad
\overline{a}=2n+1-a=\overline{i},\nonumber\\
&&a=\overline{i},\qquad \overline{a}=2n+1-a=i,\qquad
\left(i=1,\cdots,n\right)\end{eqnarray} Similarly
$\left(b,c\right)$ can be $\left(j,\overline{j}\right)$,
$\left(k,\overline{k}\right)$ respectively over the same domain.
The notations (2.8), (2.9) make the formalism much more compact.

Using (2.3-9) one obtains
\begin{equation}
\widehat{\mathrm{B}}\left|abc\right\rangle=f_0\left|abc\right\rangle+
f_1\left|a\overline{b}\overline{c}\right\rangle+f_2\left|\overline{a}b\overline{c}\right\rangle+f_3\left|\overline{a}\overline{b}c\right\rangle
\end{equation}
with
\begin{eqnarray}
&&\left(f_0,f_1,f_2,f_3\right)=\frac
1{\sqrt{\left(1+z^2\right)\left(1+z'^2\right)\left(1+z''^2\right)}}\left(\left(1-zz'\right),\left(-1\right)^c\left(z+z'\right),\right.\nonumber\\
&&\phantom{\left(f_0,f_1,f_2,f_3\right)=}\left.\left(-1\right)^b
\left(z+z'\right),\left(-1\right)^{b+c}z''\left(z'-z\right)\right),\end{eqnarray}
Noting that
\begin{eqnarray}
&&\left(1-zz'\right)^2+\left(z+z'\right)^2=\left(1+z^2\right)\left(1+z'^2\right),\nonumber\\
&&\left(z+z'\right)^2+z''^2\left(z'-z\right)^2=z''^2\left(1+z^2\right)\left(1+z'^2\right)\end{eqnarray}
one immediately verifies that, consistently with unitarity of
$\widehat{\mathrm{B}}$,
\begin{equation}
f_0^2+f_1^2+f_2^2+f_3^2=1.
\end{equation}
On the left of (2.10), $\left|abc\right\rangle$ is by definition a
separable product of pure states, hence unentangled. One the right
the superposition can be shown to imply entanglement. On assuming
it can be expressed as a product $\left(\sum
x_i\left|x_i\right\rangle\right)\otimes \left(\sum
y_i\left|y_i\right\rangle\right)\otimes\left(\sum
z_i\left|z_i\right\rangle\right)$ one runs into contradictions. In
Sec. 3 we will go much further. We will obtain explicitly the
intrinsic 3-body entanglement (3-tangle)and the 2-body
entanglements (2-tangles) of the three subsystems. They will be
expressed in terms of $\left(f_0,f_1,f_2,f_3\right)$ of (2.11).

\paragraph{(II) Complex, even dimensional, multi-parameter unitary braid
matrices:} In a series of papers (some of which are cited in Ref.
3) we have constructed a class multiparameter braid matrices, the
number of such parameters increasing as $N^2$ with the dimension
$N^2\times N^2$ of the matrix for both even and odd dimensions
$\left(N=2,3,4,5,6,\cdots\right)$. It was then noted \cite{2,3}
that for all these parameters pure imaginary this class
corresponds to unitary braid matrices. (One may also consider real
parameters with $\theta$ imaginary.) In this subsection we
restrict our considerations to even-dimensional matrices. For $N$
odd special features arise which are best treated separately (Sec.
4). The even dimensional, unitary $\left(2n\right)^2 \times
\left(2n\right)^2$ matrix is given by
\begin{equation}
\widehat{\mathrm{R}}\left(\theta\right)=\sum_{\epsilon}\sum_{i,j}e^{m_{ij}^{(\epsilon)}\theta}\left(P_{ij}^{(\epsilon)}+P_{i\bar{j}}^{(\epsilon)}\right)
\end{equation}
the definitions of the projectors being
\begin{equation}
P_{ab}^{(\epsilon)}=\frac 12\left\{\left(aa\right)\otimes
\left(bb\right)+\left(\bar{a}\bar{a}\right)\otimes
\left(\bar{b}\bar{b}\right)+\epsilon\left[\left(a\bar{a}\right)\otimes
\left(b\bar{b}\right)+\left(\bar{a}a\right)\otimes
\left(\bar{b}b\right)\right]\right\},
\end{equation}
where $a=\left(i,\overline{i}\right)$,
$b=\left(j,\overline{j}\right)$ runs over $2n$ values and
$\overline{a}=2n+1-a$ and $\overline{b}=2n+1-b$. The explicit
forms of $\hat{R}\left(\theta\right)$ for $N=\left(2,4\right)$ are
given in Ref. 3 where the transition to unitarity is formulated in
Sec. 3 ($m_{ij}^{\left(\epsilon\right)}\longrightarrow
\mathrm{i}m_{ij}^{\left(\epsilon\right)}$, with the coefficient
$\mathrm{i}=\sqrt{-1}$).

Consider again the action of
\begin{equation}
\widehat{\mathrm{B}}=\widehat{\mathrm{R}}_{12}\left(\theta\right)\widehat{\mathrm{R}}_{23}\left(\theta+\theta'\right)
\widehat{\mathrm{R}}_{12}\left(\theta'\right).
\end{equation}
One obtains (compare (2.10), (2.11))
\begin{equation}
\widehat{\mathrm{B}}\left|abc\right\rangle=f_0\left|abc\right\rangle+
f_1\left|a\overline{b}\overline{c}\right\rangle+f_2\left|\overline{a}b\overline{c}\right\rangle+f_3\left|\overline{a}\overline{b}c\right\rangle
\end{equation}
but now with coefficients defined below. Set
\begin{equation}
\lambda_{\pm}=m_{ab}^{\left(\pm\right)}\left(\theta+\theta'\right),\qquad
\mu_{\pm}=m_{bc}^{\left(\pm\right)}\left(\theta+\theta'\right)
\end{equation}
with only the sum $\left(\theta+\theta'\right)$ as factor above
(in contrast to $\tanh\theta$, $\tanh\theta'$,
$\tanh\left(\theta+\theta'\right)$ all playing roles in the
previous case). In terms of $\left(\lambda,\mu\right)$ one has
\begin{eqnarray}
&&f_0=\frac
14\left(e^{\mathrm{i}\lambda_+}+e^{\mathrm{i}\lambda_-}\right)\left(e^{\mathrm{i}\mu_+}+e^{\mathrm{i}\mu_-}\right)\nonumber\\
&&f_1=\frac
14\left(e^{\mathrm{i}\lambda_+}+e^{\mathrm{i}\lambda_-}\right)\left(e^{\mathrm{i}\mu_+}-e^{\mathrm{i}\mu_-}\right)\nonumber\\
&&f_2=\frac
14\left(e^{\mathrm{i}\lambda_+}-e^{\mathrm{i}\lambda_-}\right)\left(e^{\mathrm{i}\mu_+}-e^{\mathrm{i}\mu_-}\right)\nonumber\\
&&f_3=\frac
14\left(e^{\mathrm{i}\lambda_+}-e^{\mathrm{i}\lambda_-}\right)\left(e^{\mathrm{i}\mu_+}+e^{\mathrm{i}\mu_-}\right)
\end{eqnarray}
satisfying the unitarity constraints
\begin{equation}
f_0f_0^++f_1f_1^++f_2f_2^++f_3f_3^+=1.
\end{equation}
Again one can easily verify that the right side of (2.17)
represents an entangled states (as for (2.10)). We will obtain
explicitly the 3-tangle and the 2-tangle in Sec. 3.

Repeated actions of $\widehat{\mathrm{B}}$ with different
parameters will modify the coefficients as
\begin{eqnarray}
&&\widehat{\mathrm{B}}'\widehat{\mathrm{B}}\left|abc\right\rangle=\widehat{\mathrm{B}}'\left(f_0\left|abc\right\rangle+
f_1\left|a\overline{b}\overline{c}\right\rangle+f_2\left|\overline{a}b\overline{c}\right\rangle+f_3\left|\overline{a}\overline{b}c\right\rangle\right)\nonumber\\
&&\phantom{\overline{\mathrm{B}}'\overline{\mathrm{B}}\left|abc\right\rangle}=g_0\left|abc\right\rangle+
g_1\left|a\overline{b}\overline{c}\right\rangle+g_2\left|\overline{a}b\overline{c}\right\rangle+g_3\left|\overline{a}\overline{b}c\right\rangle
\end{eqnarray}
with
\begin{eqnarray}
&&g_0=\left(f_0f_0'+f_1f_1'+f_2f_2'+f_3f_3'\right)\nonumber\\
&&g_1=\left(f_0f_1'+f_1f_0'+f_2f_3'+f_3f_2'\right)\nonumber\\
&&g_2=\left(f_0f_2'+f_1f_3'+f_2f_0'+f_3f_1'\right)\nonumber\\
&&g_3=\left(f_0f_3'+f_1f_2'+f_2f_1'+f_3f_0'\right).
\end{eqnarray}
Here $\left(f_0',f_1',f_2',f_3'\right)$ are coefficients due to
the action of $\widehat{\mathrm{B}}'$ alone. This set may belong
to the same class as $\left(f_0,f_1,f_2,f_3\right)$ or to the
other one of our two classes (see (2.11) and (2.19)). The process
can be repeated remaining always in the closed subspace
$\left(\left|abc\right\rangle,\left|a\overline{b}\overline{c}\right\rangle,
\left|\overline{a}b\overline{c}\right\rangle,\left|\overline{a}\overline{b}c\right\rangle\right)$.
(Starting with the same  $\left|a\right\rangle$ but with different
$\left|b'\right\rangle$, $\left|c'\right\rangle$, for example, and
continuing thus one can entangle each individual state with all
the others successively in the total base space. See the relevant
remarks in sec. 6.) The essential features of the aspects that
interest us principally (analyzed in sec. 3) can be shown to be
conserved under iterations indicated above. For that reason, and
also for simplicity, we will restrict our study (sec. 3) to the
two sets of coefficients (2.11) and (2.19).

\section{Computation of quantum entanglements}
\setcounter{equation}{0}

We now extract from (2.10), (2.11) and (2.17), (2.19) respectively
the quantum entanglements generated by $\widehat{\mathrm{B}}$
acting on the pure product state $\left|abc\right\rangle$, where
$\left|abc\right\rangle$ can be any triplet selected from the
$N^3$ dimensional base space. For spin $\frac 12$ particles
\begin{equation}
\left|a\right\rangle\in
\left(\left|+\right\rangle,\left|-\right\rangle\right)\equiv
\left(\left|1\right\rangle,\left|0\right\rangle\right)
\end{equation}
For higher spins (with $\left|\overline{a}\right\rangle$ now
written as $\left|-a\right\rangle$)
\begin{equation}
\left|a\right\rangle\in
\left(\left|N\right\rangle,\left|N-1\right\rangle,\cdots,\left|1-N\right\rangle,\left|-N\right\rangle\right)
\end{equation}
and similarly for
$\left(\left|b\right\rangle,\left|c\right\rangle\right)$. Under
the action of the braiding operator $\widehat{\mathrm{B}}$
(defined in (1.4) with
$\left|\overline{a}\right\rangle=\left|N-a+1\right\rangle\equiv
\left|-a\right\rangle$)
\begin{equation}
\widehat{\mathrm{B}}\left|abc\right\rangle=f_0\left|abc\right\rangle+
f_1\left|a\overline{b}\overline{c}\right\rangle+f_2\left|\overline{a}b\overline{c}\right\rangle+f_3\left|\overline{a}\overline{b}c\right\rangle,
\end{equation}
where $\left(f_0,f_1,f_2,f_3\right)$ are given by (2.11) and
(2.19) for our two classes respectively. For spin $\frac 12$ one
has two subsets
\begin{equation}
\left(\left|111\right\rangle,\left|100\right\rangle,\left|010\right\rangle,\left|001\right\rangle\right)\qquad
\hbox{and}\qquad
\left(\left|000\right\rangle,\left|011\right\rangle,\left|101\right\rangle,\left|110\right\rangle\right)\qquad
\end{equation}
For higher spin, as noted before, one can start with the same
$\left|a\right\rangle$ but
$\left(\left|b\right\rangle,\left|c\right\rangle\right)$ chosen
from all the other possibilities. But for each initial choice one
remains, under the action $\widehat{\mathrm{B}}$, in the subspace
given by the right hand of (3.3). This is the very special,
fundamental, property of our unitary matrices. This allows us,
even for higher spins, to implement systematically the formalism
and concepts of Coffman, Kundu and Wootters (CKW) concerning
3-particle entanglements (and corresponding 2-particle ones for
the three subsystems) in Ref. 4. This we now proceed to do.

The density matrix corresponding (3.3) is
\begin{eqnarray}
&&\rho_{123}=\left(f_0\left|abc\right\rangle+
f_1\left|a\overline{b}\overline{c}\right\rangle+f_2\left|\overline{a}b\overline{c}\right\rangle+f_3\left|\overline{a}\overline{b}c\right\rangle\right)
\left(f_0^+\left\langle cba\right|+\right.\nonumber\\
&&\phantom{\rho_{123}=}\left.
f_1^+\left\langle\overline{c}\overline{b}a\right|+f_2^+\left\langle\overline{c}b\overline{a}\right|+
f_3^+\left\langle c\overline{b}\overline{a}\right|\right)
\end{eqnarray}
Tracing out $c$ one obtains
\begin{eqnarray}
&&\rho_{12}=f_0f_0^+\left|ab\right\rangle\left\langle ba\right|+
f_0f_3^+\left|ab\right\rangle\left\langle
\overline{b}\overline{a}\right|+f_1f_1^+\left|a\overline{b}\right\rangle\left\langle
\overline{b}a\right|+f_1f_2^+\left|a\overline{b}\right\rangle\left\langle
b\overline{a}\right|+\nonumber\\
&&\phantom{\rho_{12}=}f_2f_2^+\left|\overline{a}b\right\rangle\left\langle
b\overline{a}\right|+
f_2f_1^+\left|\overline{a}b\right\rangle\left\langle
\overline{b}a\right|+f_3f_3^+\left|\overline{a}\overline{b}\right\rangle\left\langle
\overline{b}\overline{a}\right|+
f_3f_0^+\left|\overline{a}\overline{b}\right\rangle\left\langle
ba\right|
\end{eqnarray}
Tracing out $b$ in $\rho_{12}$ one obtains  a diagonal
\begin{eqnarray}
&&\rho_{1}=\left(f_0f_0^++f_1f_1^+\right)\left|a\right\rangle\left\langle
a\right|+\left(f_2f_2^++f_3f_3^+\right)\left|\overline{a}\right\rangle\left\langle
\overline{a}\right|
\end{eqnarray}
One can write down $\left(\rho_{13},\rho_{13}\right)$,
$\left(\rho_{2},\rho_{3}\right)$ from symmetry. Thus, for example,
\begin{eqnarray}
&&\rho_{2}=\left(f_0f_0^++f_2f_2^+\right)\left|b\right\rangle\left\langle
b\right|+\left(f_3f_3^++f_1f_1^+\right)\left|\overline{b}\right\rangle\left\langle
\overline{b}\right|
\end{eqnarray}
and so on.

The spin-flipped matrix
\begin{eqnarray}
&&\widetilde{\rho}_{AB}=\begin{vmatrix}
  0 & -\mathrm{i} \\
  \mathrm{i} & 0 \\
\end{vmatrix}\otimes \begin{vmatrix}
  0 & -\mathrm{i} \\
  \mathrm{i} & 0 \\
\end{vmatrix}\rho_{AB}^{+}\begin{vmatrix}
  0 & -\mathrm{i} \\
  \mathrm{i} & 0 \\
\end{vmatrix}\otimes \begin{vmatrix}
  0 & -\mathrm{i} \\
  \mathrm{i} & 0 \\
\end{vmatrix}
\end{eqnarray}
can now be obtained for
$\left(\rho_{12},\rho_{23},\rho_{13}\right)$ and then the products
$\left(\rho_{AB}\widetilde{\rho}_{AB}\right)$. Thus
\begin{eqnarray}
&&\rho_{12}\widetilde{\rho}_{12}=2\begin{vmatrix}
  f_0f_0^+f_3f_3^+ & 0 & 0 & f_0^2f_0^+f_3^+ \\
  0 & f_1f_1^+f_2f_2^+ & f_1^2f_1^+f_2^+ & 0 \\
0 & f_2^2f_1^+f_2^+ & f_1f_1^+f_2f_2^+ & 0 \\
f_3^2f_0^+f_3^+ & 0 & 0 & f_0f_0^+f_3f_3^+ \\
\end{vmatrix}
\end{eqnarray}
The products $\left(\rho_{13}\widetilde{\rho}_{13}\right)$,
$\left(\rho_{23}\widetilde{\rho}_{23}\right)$ are related to the
result above through evident permutations of the indices
$\left(1,2,3\right)$. The eigenstates can be read off as
$\left|\begin{matrix}
  f_0/f_3\\
  0 \\
  0 \\
  \pm 1\\
\end{matrix}\right\rangle$, $\left|\begin{matrix}
  0\\
  f_1/f_2\\
  \pm 1 \\
  0 \\
  \end{matrix}\right\rangle$.
The eigenvalues of $\left(\rho_{12}\widetilde{\rho}_{12}\right)$
are
\begin{equation}
\left(\lambda_1^2,\lambda_2^2,\lambda_3^2,\lambda_4^2\right)=
4\left(f_0f_0^+f_3f_3^+,f_1f_1^+f_2f_2^+,0,0\right)\end{equation}
(The ordering of the first two roots depends on values of the
parameters in $\widehat{\mathrm{B}}$.)

Taking square roots the "concurrence" is
\begin{equation}
C_{12}=2\left|\left(f_0f_0^+f_3f_3^+\right)^{1/2}-\left(f_1f_1^+f_2f_2^+\right)^{1/2}\right|
\end{equation}
Similarly
\begin{eqnarray}
&&C_{23}=2\left|\left(f_0f_0^+f_1f_1^+\right)^{1/2}-\left(f_2f_2^+f_3f_3^+\right)^{1/2}\right|\\
&&C_{13}=2\left|\left(f_0f_0^+f_2f_2^+\right)^{1/2}-\left(f_1f_1^+f_3f_3^+\right)^{1/2}\right|
\end{eqnarray}
Note that the product $\left(\lambda_1\lambda_2\right)$ is the
same for the three subsystems. Thus from (17) and (24) of CKW, the
3-tangle is
\begin{equation}
\tau_{123}=16\left(f_0f_0^+f_1f_1^+f_2f_2^+f_3f_3^+\right)^{1/2}.
\end{equation}
The invariance of $\tau_{123}$ under permutations of the particles
$(1,2,3)$ (i.e. $\left(a,b,c\right)$) is evident above. Having
obtained the results in terms of $\left(f_0,f_1,f_2,f_3\right)$ we
proceed below to study them for our two classes implementing
(2.11) and (2.19). We start with $\tau_{123}$ since the crucial
role of $\widehat{\mathrm{B}}$ is to entangle 3 particles.

\paragraph{(I):} From (2.11) for real $\left(\theta,\theta'\right)$
\begin{equation}
\tau_{123}=16\left(f_0f_0^+f_1f_1^+f_2f_2^+f_3f_3^+\right)^{1/2}=16\left(f_0f_1f_2f_3\right)=
16\frac{(1-zz')(z+z')^2z''|z-z'|}{((1+z^2)(1+z'^2)(1+z''^2))^2}
\end{equation}
Here $z=\tanh\theta$, $z'=\tanh\theta'$,
$z''=\tanh\left(\theta+\theta'\right)$ and $1\geq
\left(z,z',z''\right)\geq 0$. Special points:

\begin{enumerate}

\item $\left(z=z'\right)$: The difference  $\left|z-z'\right|$ in (3.16) arises from
anticommutativity of $\left(J,K\right)$ in (2.4). For $z=z'$,
$f_3=0$ and hence $\tau_{123} =0$ in (3.15). From (3.12), (3.13),
(3.14) one has non-zero 2-tangles. We exclude this point, being
particulary interested in 3-tangle.

\item $\left(z=-z'\right)$: Now $z''=0$, i.e. $\left(\theta+\theta'\right)=0$. This is a trivial point. Now in
(1.4)
\begin{equation}
\widehat{\mathrm{B}}=\widehat{\mathrm{R}}_{12}\left(\theta\right)\widehat{\mathrm{R}}_{23}\left(0\right)\widehat{\mathrm{R}}_{12}\left(-\theta\right)=
\widehat{\mathrm{R}}_{23}\left(-\theta\right)\widehat{\mathrm{R}}_{12}\left(0\right)\widehat{\mathrm{R}}_{23}\left(\theta\right),
\end{equation}
where
\begin{equation}
\left(\widehat{\mathrm{R}}_{23}\left(0\right),\widehat{\mathrm{R}}_{12}\left(0\right)\right)=I\otimes
I\otimes I\end{equation} and due to unitarity
\begin{equation}
\widehat{\mathrm{R}}_{12}\left(\theta\right)\widehat{\mathrm{R}}_{12}\left(-\theta\right)=\widehat{\mathrm{R}}_{23}\left(\theta\right)
\widehat{\mathrm{R}}_{23}\left(-\theta\right)=I\otimes I\otimes I.
\end{equation}
We exclude also this point.

\item $\left(z=1,z'=0\right)$, $\left(z=0,z'=1\right)$: For both points
\begin{equation}
z''=1
\end{equation}
These limiting cases provide the maximal value $\tau_{123}=1$.
They correspond to
\begin{eqnarray}
&&f_0=f_1=f_2=f_3=\frac 12,\\
&&B\left|abc\right\rangle=\frac 12\left(\left|abc\right\rangle+
\left|a\overline{b}\overline{c}\right\rangle+\left|\overline{a}b\overline{c}\right\rangle+\left|\overline{a}\overline{b}c\right\rangle\right).
\end{eqnarray}
The overall factor $\frac 12$ contributes $\frac 1{2^4}$ to cancel
exactly the factor 16 in (3.16), which arose from the fact that
our $\widehat{\mathrm{B}}$ acting on a product state
$\left|abc\right\rangle$ gives a superposition of 4 states leading
to entanglements.
\end{enumerate}

Away from such points, for the generic case, keeping in mind
(2.12) we define
\begin{equation}
x=\frac{\left(1-zz'\right)}{\left(z+z'\right)},\qquad
y=\frac{z''\left|z-z'\right|}{\left(z+z'\right)}
\end{equation}
and write
\begin{equation}
\tau_{123}=\frac{16xy}{\left(x^2+y^2+2\right)^2}<\frac{4xy}{\left(xy+1\right)^2}
=\left(\frac{\left(\left(xy\right)^{1/2}+\left(xy\right)^{-1/2}\right)}{2}\right)^{-2}<1
\end{equation}
From (2.11), (3.12), (3.13), (3.14) the 2-particle concurrences
are
\begin{equation}
C_{12}=2\left|f_0f_3-f_1f_2\right|=\frac{4z''z'}{\left(1+z'^2\right)\left(1+z''^2\right)},\qquad(\hbox{for}\,\,\left|z-z'\right|=z-z')
\end{equation}
and
\begin{equation}
C_{12}=\frac{4z''z}{\left(1+z^2\right)\left(1+z''^2\right)},\qquad(\hbox{for}\,\,\left|z-z'\right|=z'-z) \\
\end{equation}
\begin{equation}
C_{23}=C_{13}=2\left|f_1\left(f_0-f_3\right)\right|=\frac{2(z+z')(1-z^2)}{\left(1+z^2\right)\left(1+z''^2\right)\left(1+zz'\right)},
\qquad(\hbox{for}\,\,\left|z-z'\right|=z-z')
\end{equation}
and
\begin{equation}
C_{23}=C_{13}=\frac{2(z+z')(1-z'^2)}{\left(1+z'^2\right)\left(1+z''^2\right)\left(1+zz'\right)},
\qquad(\hbox{for}\,\,\left|z-z'\right|=z'-z)
\end{equation}
For $\left(z=1,z'=0,z''=1\right)$ and also for
$\left(z=0,z'=1,z''=1\right)$ $C_{12}=C_{23}=C_{13}=0$, while
$\tau_{123}=1$ attaining the maximum. It is the situation one
finds in GHZ state though that is quite different otherwise. (We
refer to the comments below (24) of CKW). One can also compare a
Borromean ring (Ref. 1, sec. 8.3, for example). If any one of the
three entangled braids is cut, the remaining two fall apart-they
are no longer entangled.

\paragraph{(II):} We now obtain the quantum entanglements induced by the
action of $\widehat{\mathrm{B}}$ in a product state
$\left|abc\right\rangle$ for our complex, multiparameter, unitary
braid operator. From (2.19),
\begin{eqnarray}
&&f_0f_0^+=\left(\cos\left(\frac{\lambda_+-\lambda_-}2\right)\cos\left(\frac{\mu_+-\mu_-}2\right)\right)^2\nonumber\\
&&f_1f_1^+=\left(\cos\left(\frac{\lambda_+-\lambda_-}2\right)\sin\left(\frac{\mu_+-\mu_-}2\right)\right)^2\nonumber\\
&&f_2f_2^+=\left(\sin\left(\frac{\lambda_+-\lambda_-}2\right)\sin\left(\frac{\mu_+-\mu_-}2\right)\right)^2\nonumber\\
&&f_3f_3^+=\left(\sin\left(\frac{\lambda_+-\lambda_-}2\right)\cos\left(\frac{\mu_+-\mu_-}2\right)\right)^2
\end{eqnarray}
satisfying
\begin{equation}
f_0f_0^++f_1f_1^++f_2f_2^++f_3f_3^+=1.
\end{equation}
Here, from (2.18)
\begin{equation}
\lambda_+-\lambda_-=\left(m_{ab}^{(+)}-m_{ab}^{(-)}\right)\left(\theta+\theta'\right),\qquad
\mu_+-\mu_-=\left(m_{bc}^{(+)}-m_{bc}^{(-)}\right)\left(\theta+\theta'\right).
\end{equation}
The 3-tangle is now from (3.15), along with (3.31),
\begin{equation}
\tau_{123}=16\left(f_0f_0^+f_1f_1^+f_2f_2^+f_3f_3^+\right)^{1/2}=
\left(\sin\left(\lambda_+-\lambda_-\right)\sin\left(\mu_+-\mu_-\right)\right)^2.
\end{equation}
From (3.12), (3.13), (3.14) the 2-tangles are
\begin{eqnarray}
&&C_{12}=\left|\sin\left(\lambda_+-\lambda_-\right)\cos\left(\mu_+-\mu_-\right)\right|,\\
&&C_{23}=\left|\cos\left(\lambda_+-\lambda_-\right)\sin\left(\mu_+-\mu_-\right)\right|,\\
&&C_{13}=0.
\end{eqnarray}

Let us take a closer look at these results.

\begin{enumerate}
\item The vanishing of $C_{13}$ is related to the fact that in the action of
$\widehat{\mathrm{B}}$ on $\left|abc\right\rangle$, the terms
involving $m_{ab}^{(\pm)}$ act on $\left|ab\right\rangle$ and
those involving $m_{bc}^{(\pm)}$ on $\left|bc\right\rangle$. Thus
$\left|b\right\rangle$ is acted on by both parts while
$\left|a\right\rangle$ and $\left|c\right\rangle$ are decoupled in
the above sense. One can alter the actions on them independently
by varying the two sets of parameters. One the other hand the
presence of $\left|b\right\rangle$ generates a coupling with
$\left|a\right\rangle$ on one hand and with $\left|c\right\rangle$
on the other. A parallel feature was absent for our class (I).
There apart from $\left(\theta,\theta'\right)$ there are no free
parameters (like the $m$'s for class (II)). And
$z''=\left(z+z'\right)\left(1+zz'\right)^{-1}$ combines
$\left(z,z'\right)$ nonlinearly.

\item Here the presence of the sum $\left(\theta+\theta'\right)$ in (3.31) makes zero
entanglements for $\left(\theta+\theta'\right)=0$ evident .Compare
the discussion related to (3.17)-(3.19).

\item The domain $0\leq \tau_{123}\leq 1$ is evident from (3.32). Compare the
discussion leading to (3.24).

\item If the ratio (for +1 or -1 below)
\begin{equation}
\left(\frac{m_{ab}^{(+)}-m_{ab}^{(-)}}{m_{bc}^{(+)}-m_{bc}^{(-)}}\right)^{\pm
1}=1,3,5,\cdots
\end{equation}
an odd integer the upper limit $\left(\tau_{123}=1\right)$ is
attained periodically in the space of rapidities as the sum
$(\theta+\theta')$ is varied. If the ratio on the right is
incommensurable one can have quasi-periodicity. Varying the
parameters $\left(m_{ab}^{(\pm)},m_{bc}^{(\pm)}\right)$ one can
sweep through different possibilities.

\item For spin $\frac 12$ ($\widehat{\mathrm{R}}$ a $4\times 4$ matrix) there is only one set
$m_{11}^{(\pm)}$
$\left(m_{bc}^{(\pm)}=m_{ab}^{(\pm)}=m_{11}^{(\pm)}\right)$. Hence
in (3.32)
\begin{equation}
\tau_{123}=\left(\sin\left(\left(m_{11}^{(+)}-m_{11}^{(-)}\right)\left(\theta+\theta'\right)\right)\right)^4
\end{equation}
This is always periodic in $\left(\theta+\theta'\right)$.
\end{enumerate}

\section{Odd dimensions}
\setcounter{equation}{0}

The real matrices (2.1-3), our class (I), have no odd dimensional
counterparts. The complex, multiparameter matrices (2.14), (2.15)
are not thus restricted. In fact the odd dimensional sequences
based on "nested sequences of projectors" were the first to be
constructed. The lowest odd dimensional $\left(9\times 9\right)$
case, with imaginary parameters for unitarity, is explicitly
presented in sec 11 of Ref. 2. The crucial difference with $N=2n$
is that for $N=\left(2n-1\right)$, $\left(n=2,3,\cdots\right)$
$\overline{i}=(2n-1)-i+1=2n-i$ and hence
\begin{equation}
\overline{n}=2n-n=n.
\end{equation}
For $N=2n$, $\overline{i}\neq i$ for each $i$. Correspondingly in
(2.15), now (for $\left(a,b\right)\neq n$)
\begin{eqnarray}
&&P_{an}^{(\epsilon)}=P_{a\overline{n}}^{(\epsilon)}=\frac
12\left\{\left(aa\right)+\left(\bar{a}\bar{a}\right)+\epsilon\left[\left(a\bar{a}\right)+\left(\bar{a}a\right)\right]\right\}\otimes
\left(nn\right),\\
&&P_{nb}^{(\epsilon)}=\frac 12 \left(nn\right)\otimes
\left\{\left(bb\right)+\left(\bar{b}\bar{b}\right)+\epsilon\left[\left(b\bar{b}\right)+\left(\bar{b}b\right)
\right]\right\},\\
&&P_{nn}=\left(nn\right)\otimes \left(nn\right),
\end{eqnarray}
If in (2.17)
\begin{equation}
\left(a,b,c\right)\neq n
\end{equation}
the results of sec. 3 can be formally carried over unchanged. But
for
\begin{equation}
\left|abc\right\rangle=\left(\left|nbc\right\rangle,\left|anc\right\rangle,\left|abn\right\rangle,
\left|nnc\right\rangle,\left|nbn\right\rangle,\left|ann\right\rangle,\left|nnn\right\rangle\right)
\end{equation}
fairly evident modifications are necessary. Some indications are
given in Ref. 2. Odd dimensional $\widehat{\mathrm{R}}$ is
necessary in dealing with particles of integer spins. If in a
3-photon state each one is in a state of polarization
$\left|\pm\right\rangle$ (i.e. $\left|\pm 1\right\rangle$), there
being no $\left|0\right\rangle$ states the results of sec. 2 (II)
can be used.

Consider now the action of $\widehat{\mathrm{B}}$ on the states
(4.6). From (4.2), (4.3), (4.4)
\begin{eqnarray}
&&P_{nb}^{(\epsilon)}\left|nb\right\rangle=\frac 12
\left|n\right\rangle\otimes
\left(\left|b\right\rangle+\epsilon\left|\overline{b}\right\rangle\right)=
\frac 12
\left(\left|nb\right\rangle+\epsilon\left|n\overline{b}\right\rangle\right)\\
&&P_{an}^{(\epsilon)}\left|an\right\rangle=\frac 12
\left(\left|a\right\rangle+\epsilon\left|\overline{a}\right\rangle\right)\otimes
\left|n\right\rangle= \frac 12
\left(\left|an\right\rangle+\epsilon\left|\overline{a}n\right\rangle\right)\\
&&P_{nn}^{(\epsilon)}\left|nn\right\rangle=\left|nn\right\rangle.
\end{eqnarray}
The action of $\widehat{\mathrm{B}}$ on the states (4.6) can now
be studied. From our point of view (links with quantum
entanglements) not only $\left|nnn\right\rangle$ but also
($\left|nnc\right\rangle$,$\left|nbn\right\rangle$,$\left|ann\right\rangle$)
are trivial since one obtains under action of
$\widehat{\mathrm{B}}$ superposition of states
$\left|nn\right\rangle\otimes
\left(\left|c\right\rangle,\left|\overline{c}\right\rangle\right)$
and so on. Only one spin is affected. Entanglement is not
produced. The states
$\left(\left|nbc\right\rangle,\left|abn\right\rangle\right)$ can
also be set aside, under the action of $\widehat{\mathrm{B}}$ the
state $\left|n\right\rangle$ remains a bystander to 2-particle
entanglements of $\left|bc\right\rangle$ and
$\left|ab\right\rangle$.

The state $\left|anc\right\rangle$ deserves a closer look. One
obtains
\begin{equation}
\widehat{\mathrm{B}}\left|anc\right\rangle=f_0
\left|anc\right\rangle+f_1\left|an\overline{c}\right\rangle+f_2\left|\overline{a}n\overline{c}\right\rangle+f_3
\left|\overline{a}nc\right\rangle.
\end{equation}
The coefficients $\left(f_0,f_1,f_2,f_3\right)$ are obtained by
setting, in (2.18), (2.19),
\begin{equation}
\lambda_{\pm}=m_{an}^{(\pm)}\left(\theta+\theta'\right), \qquad
\mu_{\pm}=m_{nc}^{(\pm)}\left(\theta+\theta'\right),
\end{equation}
But now tracing out indices (with $\overline{n}=n$) leads to
differences. As compared to (3.6), now (with $b=\overline{b}=n$)
\begin{eqnarray}
&&\rho_{12}=\left(f_0f_0^++f_1f_1^+\right)\left|an\right\rangle\left\langle
na\right|+
\left(f_0f_3^++f_1f_2^+\right)\left|an\right\rangle\left\langle
n\overline{a}\right|+\nonumber\\
&&\phantom{\rho_{12}=}\left(f_3f_0^++f_2f_1^+\right)\left|\overline{a}n\right\rangle\left\langle
na\right|+\left(f_2f_2^++f_3f_3^+\right)\left|\overline{a}n\right\rangle\left\langle
n\overline{a}\right|.
\end{eqnarray}
One has $2\times 2$ matrix now. One the other hand tracing out $n$
in $\rho_{12}$
\begin{eqnarray}
&&\rho_{1}=\left(f_0f_0^++f_1f_1^+\right)\left|a\right\rangle\left\langle
a\right|+
\left(f_0f_3^++f_1f_2^+\right)\left|a\right\rangle\left\langle
\overline{a}\right|+\nonumber\\
&&\phantom{\rho_{12}=}\left(f_3f_0^++f_2f_1^+\right)\left|\overline{a}\right\rangle\left\langle
a\right|+
\left(f_2f_2^++f_3f_3^+\right)\left|\overline{a}\right\rangle\left\langle
\overline{a}\right|.
\end{eqnarray}
This is no longer diagonal like (3.7).

We will not analyze such cases further in this paper. For $N=2n$,
at the centre of the matrix $\widehat{\mathrm{R}}$ is the square
lattice with corners, which can be denoted as
$\left(nn,n\overline{n},\overline{n}n, \overline{n}
\overline{n}\right)$. For  $N=2n-1$, (since $\overline{n}=n$) this
reduces to the point $\left(nn\right)$ common to the diagonal and
the anti-diagonal. This is the source of difference. When this
common point is not involved in $\left|abc\right\rangle$ the
results correspond for even and odd dimensions.

\section{Entanglement via a special coupling of 3 spins}
\setcounter{equation}{0}

This section is a brief digression. We restrict our remarks here
to 3 spin $\frac 12$ particles. For this case, in the study of
entanglements, basic roles are usually attributed to the states
\begin{eqnarray}
&&\left|GHZ\right\rangle=\frac 1{\sqrt{2}}
\left(\left|000\right\rangle+\left|111\right\rangle\right)\\
&&\left|W\right\rangle=\frac 1{\sqrt{3}}
\left(\left|001\right\rangle+\left|010\right\rangle+\left|100\right\rangle\right)\\
&&\left|\widetilde{W}\right\rangle=\frac 1{\sqrt{3}}
\left(\left|110\right\rangle+\left|101\right\rangle+\left|011\right\rangle\right)
\end{eqnarray}
(See Ref. 7 and sources cited there.) Their local unitary
transformations can also be considered \cite{5}.

Our approach via the braiding operator $\widehat{\mathrm{B}}$ led
to states of the type
$\left(f_0\left|000\right\rangle+f_1\left|011\right\rangle+\right.$
$\left.f_2\left|101\right\rangle+f_3\left|110\right\rangle\right)$
and
$\left(g_0\left|111\right\rangle+g_1\left|100\right\rangle+g_2\left|010\right\rangle+g_3\left|001\right\rangle\right)$
with normalized coefficients (with parameter dependence)
\begin{eqnarray}
&&f_0f_0^++f_1f_1^++f_2f_2^++f_3f_3^+=1,\\
&&g_0g_0^++g_1g_1^++g_2g_2^++g_3g_3^+=1
\end{eqnarray}
given in sec 2. The states  $\left|000\right\rangle$ and
$\left|111\right\rangle$ of $\left|GHZ\right\rangle$ are
separately superposed respectively with those of
$\left|\widetilde{W}\right\rangle$, $\left|W\right\rangle$
respectively and these 4-term in the superpositions have been
thoroughly studied in the preceding sections. Though we are
particularly interested in such cases it is interesting to point
out the direct relations of $\left|GHZ\right\rangle$,
$\left|\widetilde{W}\right\rangle$ and $\left|W\right\rangle$ to a
coupling of 3 angular momenta
$\left(\overrightarrow{\mathbf{J}}_1,\overrightarrow{\mathbf{J}}_2,
\overrightarrow{\mathbf{J}}_3\right)$ to obtain eigenstates of
\begin{equation}
Z=\left(\overrightarrow{\mathbf{J}}_1\times\overrightarrow{\mathbf{J}}_2\right)\cdot
\overrightarrow{\mathbf{J}}_3.
\end{equation}
These also, of course, eigenstates of
\begin{equation}
\overrightarrow{\mathbf{J}}^2=\overrightarrow{\mathbf{J}}_1^2+\overrightarrow{\mathbf{J}}_2^2+
\overrightarrow{\mathbf{J}}_3^2
\end{equation}
and
\begin{equation}
J^0=J_1^0+J_2^0+J_3^0.
\end{equation}
$J_0$ being the third component in the circular ones $\left(J_+,
J_-, J_0\right)$. Such a coupling was proposed by A. Chakrabarti
long ago \cite{8}. It was also proposed by J.M. L\'evy-Leblond and
M. Nahas \cite{9}. From all the results of Ref. 8, concerning
states $\left|jm\zeta\right\rangle$ $\left(j(j+1),m,\zeta\right)$
denoting eigenvalues of ($\left(\overrightarrow{\mathbf{J}}^2,
J^0,Z\right)$ respectively) we mention one. For the maximal value
$j=j_1+j_2+j_3$ always $\zeta=0$. For $j_1=j_2=j_3=\frac 12$ and
$j=\frac 32$,
\begin{eqnarray}
&&\frac 1{\sqrt{2}}\left(\left|\frac 32\,\frac 32\,\,
0\right\rangle\pm \left|\frac 32 \,-\frac 32\,\,
0\right\rangle\right)=\frac
1{\sqrt{2}}\left(\left|000\right\rangle\pm
\left|111\right\rangle\right),\\
&&\left|\frac 32\,\frac 12\,\, 0\right\rangle=\left|W\right\rangle,\\
&&\left|\frac 32\, -\frac 12\,\,
0\right\rangle=\left|\widetilde{W}\right\rangle.
\end{eqnarray}
(The  states on the right, unlike those on the left, are those of
(5.1), (5.2), (5.3).) For $j=\frac 12$ one has non zero $\zeta$.
From (A.13) of Ref. 8 (with $\zeta=\pm \frac{\sqrt{3}}4$)
\begin{eqnarray}
&&\left|\frac 12\,\frac 12\,
\pm\frac{\sqrt{3}}4\right\rangle=\frac 1{\sqrt{3}}\left(e^{\pm
\mathrm{i}\frac{2\pi}3}\left|011\right\rangle+ e^{\mp
\mathrm{i}\frac{2\pi}3}\left|101\right\rangle+\left|110\right\rangle\right),\\
&&\left|\frac 12\,-\frac 12\,
\pm\frac{\sqrt{3}}4\right\rangle=\frac 1{\sqrt{3}}\left(e^{\mp
\mathrm{i}\frac{2\pi}3}\left|100\right\rangle+ e^{\pm
\mathrm{i}\frac{2\pi}3}\left|010\right\rangle+\left|001\right\rangle\right),
\end{eqnarray}
We conclude by noting:

\begin{enumerate}
\item This coupling was proposed due to its symmetries under the
permutations of the spins $\left(\overrightarrow{\mathbf{J}}_1,
\overrightarrow{\mathbf{J}}_2,\overrightarrow{\mathbf{J}}_3\right)$.
Such symmetries are notoriously lacking for the standard 2-step
coupling via C.G. coefficients where permutations are related to
6-$j$ symbols. One has to label the intermediate step additionally
with $(j_1,j_2)$ or $(j_1,j_3)$ or $(j_2,j_3)$. Reduction under
the rotation group implementing $Z$ gives simultaneous reduction,
without additional effort, under $S_3$ the group of permutations
of the 3 particles .

\item Here are see how this formalism leads directly to states famous in the study of quantum
entanglements.

\end{enumerate}

\section{Discussions}
\setcounter{equation}{0}

We want to emphasize how in larger dimensions each object (say,
spin states of component particles) is seen to be entangled with
all the others through a full exploitation of our formalism. Since
our approach is via the braiding operator $\widehat{\mathrm{B}}$
(defined in (1.4)) we start by picking out a triplet
\begin{equation}
\left|a\right\rangle\otimes\left|b\right\rangle\otimes\left|c\right\rangle
\equiv \left|abc\right\rangle,
\end{equation}
where $\left(a,b,c\right)$ is any element among the basis states
spanning $\mathrm{V}_N\otimes \mathrm{V}_N\otimes \mathrm{V}_N$
and obtain the entangled superpositions studied (sec. 3)
\begin{equation}
\widehat{\mathrm{B}}\left|abc\right\rangle=f_0
\left|abc\right\rangle+f_1\left|a\overline{b}\overline{c}\right\rangle+f_2\left|\overline{a}b\overline{c}\right\rangle+f_3
\left|\overline{a}\overline{b}c\right\rangle.
\end{equation}
where $\left|\overline{a}\right\rangle=\left|N-a+1\right\rangle$
and so on. The states $\left(\left|a\right\rangle,
\left|\overline{a}\right\rangle\right)$,
$\left(\left|b\right\rangle,
\left|\overline{b}\right\rangle\right)$,
$\left(\left|c\right\rangle,
\left|\overline{c}\right\rangle\right)$, of the subsystems are
involved above. But now one can start again with any triplet
$\left|ab'c'\right\rangle$, $\left|a''bc''\right\rangle,\cdots$,
($b'\neq b$ or $\overline{b}$ and so on) covering thus
systematically all possible choices in $\mathrm{V}_N\otimes
\mathrm{V}_N\otimes \mathrm{V}_N$ and then implement the action of
$\widehat{\mathrm{B}}$. Thus at the end each
$\left|a\right\rangle$ will be entangled with each
$\left|b\right\rangle$ and each $\left|c\right\rangle$. At each
step a quadruplet $\left(\left|abc\right\rangle,
\left|a\overline{b}\overline{c}\right\rangle,
\left|\overline{a}b\overline{c}\right\rangle,
\left|\overline{a}\overline{b}c\right\rangle\right)$ will be
involved, this being the essential property of both classes of
unitary braid matrices we propose (with non-zero terms on the
diagonal and the anti-diagonal only).

Consider the simplest non-trivial case. For three spin half
particles the two quadruplets will be
$\left(\left|000\right\rangle, \left|011\right\rangle,
\left|101\right\rangle,\left|110\right\rangle\right)$
$\left(\left|111\right\rangle, \left|100\right\rangle,
\left|010\right\rangle, \left|001\right\rangle\right)$. But
already it is evident that starting by turns with, say
$\left(\left|000\right\rangle, \left|001\right\rangle\right)$
finally $\left|a\right\rangle= \left|0\right\rangle$ will be
entangled with $\left|bc\right\rangle=\left(
\left|00\right\rangle, \left|01\right\rangle,
\left|10\right\rangle, \left|11\right\rangle\right)$ i.e. with all
possible states of $\left|b\right\rangle$ and
$\left|c\right\rangle$. This is a general feature.

In sec .5 we have contrasted our typical superposition (6.2), to
the prominent roles in the study of 3-particle entanglements of
the states $\left(\left|GHZ\right\rangle,
\left|W\right\rangle,\left|\widetilde{W}\right\rangle\right)$
given in (5.1)-(5.3). It is implicit in our formalism that the
maximum 3-tangle is obtained for
\begin{equation}
f_0=f_1=f_2=f_3=\frac 12
\end{equation}
namely $\frac
1{2}\left(\left|111\right\rangle+\left|100\right\rangle+
\left|010\right\rangle+\left|001\right\rangle\right)$ and $\frac
1{2}\left(\left|000\right\rangle+\left|011\right\rangle+
\left|101\right\rangle+\left|110\right\rangle\right)$. The
$\left(d_1,d_2,d_3\right)$ defined in (21) of CKW \cite{4} are for
both the cases above
\begin{equation}
d_1=0,\qquad d_2=0,\qquad d_3=\frac 1{16}
\end{equation}
and hence (see (22)-(24) of CKW)
\begin{equation}
\tau_{123}=4\left|d_1-2d_2+4d_3\right|=1.
\end{equation}
The value (6.4) are the same as for $\left|GHZ\right\rangle$. For
the maximal superposition (normalized sum of the quadruplets with
all coefficients equal) $\frac
1{\sqrt{8}}\left(\left|000\right\rangle+\left|011\right\rangle+
\left|101\right\rangle+\left|111\right\rangle+\right.$
$\left.\left|100\right\rangle
+\left|010\right\rangle+\left|001\right\rangle\right)$ there is a
striking change. Now,
\begin{eqnarray}
&&\left(d_1,d_2,d_3\right)=\frac 1{2^6}\left(4,6,2\right)\\
&&\tau_{123}=\frac 1{2^6}\left(4-2\cdot 6+4\cdot 2\right)=0.
\end{eqnarray}

One reaches now the lower bound. CKW  notes below (25) "It would
be very interesting to know which of the results of this paper
generalize to larger objects or to larger collections of objects".
Our formalism furnishes one possible approach to many component
objects and their collections. We have not answered the question
whether the entanglement in larger dimensions can be formulated in
a systematically hierarchical fashion, involving simultaneously
more and more objects. Our motivation has been "entangling
topological and quantum entanglements" via the braiding operator
$\widehat{\mathrm{B}}$ corresponding to third Reidemeister move.
Having constructed unitary classes $\widehat{\mathrm{B}}$ are were
able to implement them to generate quantum entanglements.

Three particle entanglements were emphasized before \cite{10}. We
have studied, for our cases, the permutation invariant measures of
entanglement of Ref. 4. The crucial feature of our treatment is
the study of $\widehat{\mathrm{B}}$ acting on $V\times V\times V$
rather than $\widehat{\mathrm{R}}$ acting on $V\times V$. The
treatment starting in sec. 2 (II) displays one aspect of the
multiple possibilities inherent in our multi-parameter models.
This can be put side by side with their role (for real parameters)
in statistical models \cite{11}. At the end of sec. 3 we briefly
evoke possible periodicity in the space of parameters. Introducing
a magnetic field (and a simple generalization of the formalism of
Ref. 12) one would obtain periodicity in time of our entangled
3-particles states. This will be studied elsewhere.

\end{document}